\documentclass[aps,pre,preprint,superscriptaddress,amsmath,amssymb]{revtex4-2}
\usepackage{graphicx}
\usepackage{dcolumn}
\usepackage{bm}
\usepackage[mathlines]{lineno}
\usepackage{epsfig}
\usepackage{epstopdf}
\usepackage{amsmath,amssymb,amsthm}
\usepackage{bm}
\usepackage{float}
\usepackage{xcolor}
\usepackage{mathtools}
\usepackage{subfig}
\usepackage{algcompatible}
\usepackage{newfloat}
\usepackage[justification=RaggedRight]{caption}
\DeclareFloatingEnvironment[
fileext=loa,
listname=List of Algorithms,
name=ALGORITHM,
placement=tbhp,
]{algorithm}

\begin{document}
\title{Effects of intermittency on turbulent transport in magnetized plasmas}
	\author{D. I. Palade}
	\email{dragos.palade@inflpr.ro}
\affiliation{National Institute of Laser, Plasma and Radiation Physics,	M\u{a}gurele, Bucharest, Romania }
	\author{L. Pomârjanschi}
\affiliation{Faculty of Physics, University of Bucharest, Romania
}	
\date{\today}

\begin{abstract}
	
We analyze how the turbulent transport of $\mathbf{E}\times \mathbf{B}$ type in magnetically confined plasmas is affected by intermittent features of turbulence. The latter are captured by the non-Gaussian distribution $P(\phi)$ of the turbulent electric potential $\phi$. Our analysis is performed at an analytical level and confirmed numerically using two statistical approaches. 
We have found that the diffusion is inhibited linearly by intermittency, mainly via the kurtosis of the distribution $P(\phi)$. The associated susceptibility for this linear process is shown to be dependent on the poloidal velocity $V_p$ and on the correlation time $\tau_c$ with a maxima at the time-of-flight $\tau_{fl}$. Intermittency does not affect the Kubo number scaling in the strong regime.


\end{abstract}


\maketitle

\section{Introduction}
\label{section_1}

Turbulence plays a major role in the dynamics and confinement of fusion plasmas, both in nowadays and future experimental devices (i.e. ITER \cite{iter2}). Low frequency instabilities evolve and saturate into a turbulent electric field $\mathbf{E}$ which, mainly via the $\mathbf{E}\times\mathbf{B}$ drift, tends to transport plasma across magnetic surfaces, toward the walls. Such radial fluxes are particularly dangerous in the SOL region \cite{Lipschultz_2007,KRASHENINNIKOV2001368} which absorbs most of the plasma exhaust and transfers it to the divertor. Understanding and controlling this type of transport in tokamak devices has been one of the major challenges for fusion science in the past decades \cite{doi:10.1063/1.2800869,doi:10.1063/1.3083299,Angioni_2009}.

Both the edge and the SOL plasma \cite{Zweben_2007} are characterized by the presence of intermittent phenomena comparable in magnitude with the amplitude of turbulence. Intermittency \cite{PhysRevLett.87.065001} is represented by transient, coherent structures with high density gradients such as blobs \cite{doi:10.1063/1.1363663,doi:10.1063/1.5086055,Cheng_2010}, Alfven modes or ELMs \cite{leonard_AW, Zohm_1996}. Regardless of its origin, intermittency leads to a non-Gaussian distribution $P(\phi)$ of the values of the turbulent electric potential $\phi(\mathbf{x},t)$. Implicitly, the departure from Gaussianity is characteristic also for field derivatives $\partial_i\phi, \partial_{ii}\phi$\cite{gonccalves2018radial}. The latter are directly related to particle's drifts and vorticity, thus, to transport.

In the present work we are concerned with understanding and describing how the non-Gaussian features of a turbulent stochastic potential $\phi(\mathbf{x},t)$ affect the transport in the case of a biased, incompressible, 2D velocity field $\mathbf{v}(\mathbf{x},t) = \hat{e}_z\times\nabla\phi(\mathbf{x},t)+\mathbf{V}_p$. This type of dynamics is relevant not only for the $\mathbf{E}\times \mathbf{B}$ drift in tokamak plasmas in the presence of a poloidal velocity $\mathbf{V}_p$, but also for other systems: incompressible fluids, astrophysical plasmas \cite{Zank_2011}, magnetic field lines wandering \cite{doi:10.1063/1.4996869,Ghilea_2011}, etc. 

Despite the amount of work done within this topic \cite{Kraichnan_1980,doi:10.1063/1.1692063,doi:10.1063/1.1699986,RevModPhys.64.961,Ottaviani_1992,PhysRevE.54.1857,PhysRevE.63.066405,PhysRevE.58.7359,PhysRevE.63.066304}, the problem of turbulent transport is, in general, poorly understood due to its complex features. The essence of the problem can be stated as it follows: we do not have simple ways to evaluate the diffusion coefficients from the Eulerian properties of the velocity field. Moreover, non-Gaussianity is rarely taken into account, by accident, when simulating realistic flows, thus, little it is known about its effects. A solid understanding of such processes might enable possibilities of controlling the turbulent transport in fusion devices and its damaging consequences.

The present theoretical analysis requires apriori knowledge of the statistical properties of the potential $\phi$: the turbulence spectrum $S(\mathbf{k},\omega)$ and the distribution $P(\phi)$. To acquire such information, one needs to use high-quality gyro-kinetic simulations \cite{Jenko_2001,doi:10.1063/1.2338775} completed by  diagnostic techniques \cite{Gao_2015,PhysRevLett.102.165005,gonccalves2018radial}. In tokamak devices, the spectrum shows a fast decay in frequency and along the radial direction with a peaked profile (at some specific wave-number $k_0$) along the poloidal direction \cite{Shafer, PhysRevLett.70.3736,doi:10.1063/1.3085792, PhysRevLett.89.225001, PhysRevLett.102.165005, Qi_2019}. One can use the spectrum  $S(\mathbf{k},\omega) = \langle |\tilde{\phi}(\mathbf{k},\omega)|^2\rangle$ to derive, as Fourier transform - under the assumption of homogeneity, the auto-correlation function $\mathcal{E}(\mathbf{x},\mathbf{x}^\prime;t,t^\prime) =\langle \phi(\mathbf{x},t)\phi(\mathbf{x}^\prime,t^\prime)\rangle \equiv \mathcal{E}(\mathbf{x}-\mathbf{x}^\prime;t-t^\prime)$. 

Regarding the PDF $P(\phi)$, the experimental evidence \cite{Riva_2019,doi:10.1063/1.1884615,gonccalves2018radial,beadle_ricci_2020,doi:10.1063/1.5100176} indicates that, in the edge and SOL regions, the potential is approximately Gaussian $P(\phi)\sim \exp{(-\phi^2)}$ at negative values $\phi<0$ and has an exponential-like distribution $P(\phi)\sim \exp{(-\lambda|\phi|)}$ in the positive range $\phi>0$. This is equivalent with a change both of the skewness and kurtosis of the distribution. Note that the departure from Gaussianity is rather the rule than the exception: all turbulence models (Navier-Stokes, Hasegawa-Mima, Vlasov-Maxwell, etc.) include convective non-linearities which lead, implicitly, to non-Gaussian solutions \cite{doi:10.1063/1.4919852,doi:10.1063/1.4984985}. 

The paper is structured as it follows. The Theory section \ref{section_2} is dedicated to a description of the model used to simulate non-Gaussian plasma turbulence. Two methods, the Decorrelation Trajectory Method (DTM) \ref{section_2.2} and the Direct Numerical Simulation (DNS) \ref{section_2.3}, used to investigate the diffusive transport are also briefly presented. The Result section \ref{section_3} is devoted to a three step analysis: semi-analytical estimations of the diffusive transport are provided \ref{section_3.1} which are further confirmed and refined by a two-level numerical analysis \ref{section_3.2} which is finally explained from an microscopic point of view \ref{section_3.3}. The last section \ref{section_4} is dedicated to conclusions and perspectives.

\section{Theory}
\label{section_2}

We describe the motion of ions in a magnetically confined plasma using a simple geometric setup: the strong magnetic field is considered constant $\mathbf{B}=B_0\hat{e}_z$ while the ions are subject to a drift-type motion in the perpendicular plane $\mathbf{x}\equiv (x,y)$ in the presence of an effective poloidal velocity (originating from magnetic drifts or plasma rotation) $\mathbf{V}_p \equiv V_p \hat{e}_y$:

\begin{align}\label{eq_1.1}
	\frac{d\mathbf{x}(t)}{dt}= \hat{e}_z\times \nabla\phi(\mathbf{x}(t),t)+ V_p \hat{e}_y
\end{align}

The statistical description of transport in this context is the following \cite{Vlad_2021, doi:10.1063/5.0035541}: an ensemble of stochastic fields $\{\phi(\mathbf{x},t)\}$ with known Eulerian properties is considered to drive an associated ensemble of trajectories via the eq. \eqref{eq_1.1}. The diffusion coefficient is computed as Lagrangian correlation $D(t) = 1/2d_t\langle \mathbf{x}^2(t)\rangle \equiv \langle \mathbf{v}(0)\mathbf{x}(t)\rangle$ with the initial conditions $\mathbf{x}(0)=0$. Using the characteristic correlation length $\lambda_c$, correlation time $\tau_c$ and velocity amplitude $V=\Phi/\lambda_c$, $\Phi = \sqrt{\langle \phi^2(\mathbf{0},0)\rangle}$, one can define the Kubo number \cite{PhysRevE.58.7359} $K_\star$:

\begin{eqnarray}\label{eq_1.2}
	K_\star = \frac{\tau_{c}}{\tau_{fl}}=\frac{V \tau_c}{\lambda_c}=\frac{\Phi \tau_c}{\lambda_c^2}
\end{eqnarray}
as a measure of the correlation time relative to the specific time-of-flight $\tau_{fl}=\lambda_c/V$. Another interpretation of $K_\star$ is that of turbulence strength. Consequently, one can distinguish two regimes of transport: quasilinear (weak, high-frequency turbulence, $K_\star\ll 1$) and the strong (low-frequency, $K_\star\gg 1$) regime. The quasi-linear asymptotic diffusion coefficient can be exactly evaluated as $D^\infty\sim K_\star^2\lambda_c^2/\tau_c$  while in the strong limit the transport is anomalous $D^\infty\sim K_\star^{1-\gamma}$ with $\gamma\in (0,1)$. Although still under debate, it has been proposed \cite{RevModPhys.64.961} and confirmed within some degree of numerical error \cite{Ottaviani_1992,PhysRevE.54.1857,hauff_2006}, that the anomalous exponent is roughly $\gamma \approx 3/10$. 

\subsection{Turbulence description}
\label{section_2.1}

The statistical approach on turbulent transport requires the modeling of the potential $\phi$ as a non-Gaussian, zero-averaged, homogeneous random field. In order to do that, we assume (as a technical commodity \cite{Palade2021,Liu2019,vio2001numerical}) that the non-Gaussian field $\phi(\mathbf{x},t)$ can be related to another, fictitious Gaussian field $\varphi(\mathbf{x},t)$ with known correlation function $\mathcal{E}(\mathbf{x}-\mathbf{x}^\prime,t-t^\prime) = \langle \varphi(\mathbf{x},t)\varphi(\mathbf{x}^\prime,t^\prime)\rangle$ via a non-linear transformation $\phi(\mathbf{x},t) = f\left(\varphi(\mathbf{x},t)\right)$. The function $f$ must be chosen such that both fields are zero-averaged $\langle \phi(\mathbf{x},t)\rangle = \langle \varphi(\mathbf{x},t)\rangle = 0$ and have the same amplitude of fluctuations $\langle \phi^2(\mathbf{x},t)\rangle = \langle \varphi^2(\mathbf{x},t)\rangle = \mathcal{E}(\mathbf{0},0)=V_0$. 

It can be easily proven that a local non-linear transformation preserves the homogeneity property. This means that the field $\phi(\mathbf{x},t)$ is also homogeneous, i.e., its correlation function is only distance dependent $\langle \phi(\mathbf{x},t)\phi(\mathbf{x}^\prime,t^\prime)\rangle =\langle f\left(\varphi(\mathbf{x},t)\right)f\left(\varphi(\mathbf{x}^\prime,t^\prime)\right)\rangle  = \mathcal{E}^\prime(\mathbf{x}-\mathbf{x}^\prime,t-t^\prime)$ 

Straightforwardly, we can generically compute the correlation of the derivatives as well as the skewness $s$ and the excess kurtosis $\delta\kappa$ of the non-Gaussian field $\phi$:
\begin{align}\label{eq_1.3}
\langle \partial_i\phi(\mathbf{x},t)\partial_j\phi(\mathbf{x}^\prime,t^\prime)\rangle & =  \langle f^\prime\left[\varphi(\mathbf{x},t)\right]f^\prime\left[\varphi(\mathbf{x}^\prime,t^\prime)\right]\partial_i\varphi(\mathbf{x},t)\partial_j\varphi(\mathbf{x}^\prime,t^\prime)\rangle\\
s = \frac{\langle \phi^3\rangle}{\langle \phi^2\rangle^{3/2}} &= \frac{\langle f^3\left[\varphi\right]\rangle}{\langle f^2\left[\varphi\right]\rangle^{3/2}}\\
\delta \kappa = \frac{\langle \phi^4\rangle}{\langle \phi^2\rangle^{2}} -3&= \frac{\langle f^4\left[\varphi\right]\rangle}{\langle f^2\left[\varphi\right]\rangle^{2}}-3
\end{align}

Experimental measurements in the SOL of various tokamak devices \cite{Riva_2019,doi:10.1063/1.871435,PhysRevLett.102.165005,doi:10.1063/1.1884615,PhysRevLett.87.065001,doi:10.1063/1.1363663} have shown PDFs of electrostatic fluctuations which exhibit longer tails as well as skewness, especially in the positive part of the distribution. For these reasons, we choose a particularly simple non-linear transformation to construct the non-Gaussian fields: $f(\varphi) \sim \varphi+\alpha \varphi^2+\beta \varphi^3-\alpha V_0$:
\begin{align}\label{eq_1.4}
	\phi &= \frac{\varphi+\alpha \varphi^2+\beta \varphi^3-\alpha V_0}{\sqrt{1+2\alpha^2V_0+3\beta V_0 (2+5\beta V_0)}}\\
	\partial_i\phi &= \frac{1+2\alpha \varphi+3\beta \varphi^2}{\sqrt{1+2\alpha^2 V_0+3\beta V_0(2+5\beta V_0)}}\partial_i\varphi\\
	s & \approx 6\alpha V_0^{1/2}\left(1+3V_0\beta\right)\\
	\delta \kappa & \approx 24\beta V_0+144 V_0^2\beta^2+48V_0\alpha^2\\
	\mathcal{E}^\prime & = \frac{\mathcal{E}  \left(2 \alpha ^2 \mathcal{E}  +6 \beta ^2 \mathcal{E} ^2+(3 \beta V_0+1)^2\right)}{1+2 \alpha ^2 V_0+3 \beta V_0 (5 \beta V_0+2)}
\end{align}
Note that, up to first order, the skewness is controlled by the $\alpha$ parameter while the kurtosis by $\beta$. Supplementary, the correlation is virtually unchanged due to its second order parametric dependence $\mathcal{E}^\prime \approx \mathcal{E} +2\alpha^2 \mathcal{E} (\mathcal{E} -V_0)+6\beta^2 \mathcal{E} \left(\mathcal{E} ^2-V_0^2\right)$. This enables us to approximate $\mathcal{E}^\prime \approx \mathcal{E}$ since $\alpha,\beta \sim 10^{-2,-1}$ for a good agreement with experimental distributions \cite{doi:10.1063/1.871435,Riva_2019}.

\subsection{Decorrelation Trajectory Method}
\label{section_2.2}

The Decorrelation Trajectory Method (DTM) has been used in the past decades to investigate various types of turbulent transport in tokamak plasmas \cite{PhysRevE.58.7359,Croitoru_2017,Vlad_2004,Vlad_2016} as well as in some astrophysical systems \cite{Negrea_2019}. The method is semi-analytical, since it describes the transport via a set of deterministic objects called Decorrelation Trajectories (DT's) \cite{PhysRevE.58.7359}. 

The main assumption of DTM is that trajectories with similar initial conditions should remain similar at all times. If this is true, one can replace the ensemble of real stochastic potentials $\{\phi(\mathbf{x},t)\}$ with a set of deterministic conditional potentials $\{\Phi^S(\mathbf{x},t)\}$ which are defined as conditional averages over real potentials in subsensembles $(S)$, i.e. $\Phi^S(\mathbf{x},t)=\langle \phi(\mathbf{x},t)\rangle^S$. The DT's $\mathbf{X}^S(t)$ are solutions for the equation \eqref{eq_1.5a} and used to compute the diffusion \eqref{eq_1.5b}: 
\begin{align}\label{eq_1.5a}
	\frac{d\mathbf{X}^S(t)}{dt} = \mathbf{V}^S(\mathbf{X}^S(t),t)=\hat{e}_z\times\nabla \Phi^S(\mathbf{X}^S(t),t) + \mathbf{V}_p\\
\label{eq_1.5b}
	D_{xx}(t)  = \langle v_x(0)x(t)\rangle\approx \int dS~P(S) V_x^S(\mathbf{0},0)X^S(t)
\end{align}

For our non-Gaussian case, we define the subsensembles via the initial values (at $\mathbf{x} = 0$ and $t=0$) of the auxiliary field $\varphi$ as: 
\begin{align}\label{eq_1.6}
	S=\{\phi(\mathbf{x},t) = f\left(\varphi(\mathbf{x},t)\right)|\partial_i\varphi(\mathbf{0},0)=\varphi_i^S; i\in \{0,x,y\}\}
\end{align}
where each $S$ has a probabilistic weight $P(S)=\prod_{i\in\{0,x,y\}}\exp\left(-(\varphi_i^S)^2/V_{ii}/2\right)$ with $V_{ii}=\langle [\partial_i\varphi(0)]^2\rangle = -\partial_{ii}\mathcal{E}(0,0)$. From a straightforward calculus of $\Phi^S = \langle\phi\rangle^S=\langle f(\varphi)\rangle^S$  we complete the DTM model: 
\begin{align}\label{eq_1.7}
\Phi^S  & = \frac{\Psi^S(1+3\beta \sigma)+\alpha(-1+\sigma +(\Psi^S)^2)+\beta (\Psi^S)^3}{\sqrt{1+2\alpha^2+6\beta+15\beta^2}}\\
\Psi^S &= \varphi_0^S\frac{\mathcal{E}(\mathbf{x},t)}{V_{0}}+\varphi_x^S\frac{\partial_x\mathcal{E}_x(\mathbf{x},t)}{V_{xx}}+\varphi_y^S\frac{\partial_y\mathcal{E}_y(\mathbf{x},t)}{V_{yy}}\\
\sigma &= V_0-\frac{\mathcal{E}^2}{V_0}-\frac{(\partial_x\mathcal{E})^2}{V_{xx}}-\frac{(\partial_y\mathcal{E})^2}{V_{yy}}
\end{align}

The function $\sigma$ is, in fact, a measure of field fluctuations within a sub-ensemble $S$ which it turns out to be independent of $S$: $\sigma (\mathbf{x},t) = \langle \delta\varphi^2(\mathbf{x},t)\rangle^S$. Note that in the Gaussian limit $\alpha=\beta= 0$ the model simplifies to $\Phi^S\to \Psi^S$, as it has been used in previous studies \cite{Croitoru_2017,Negrea_2019,PhysRevE.58.7359,Vlad_2016}. The DTM method is equivalent with the neglecting trajectory fluctuations  \cite{Vlad_2004_SS} within a subsensemble. For more details on the method see \cite{Vlad_2004,PhysRevE.58.7359,PhysRevE.63.066304,Croitoru_2017,Vlad_2004}.

\subsection{Direct numerical simulation method}
\label{section_2.3}

The purpose of direct numerical simulations (DNS) is to investigate the turbulent transport as it is, without resorting to any approximations, closures, or supplementary models. 

In our case, this is achieved constructing a statistical ensemble of Gaussian random fields (GRFs) $\varphi(\mathbf{x},t)$ with the correct correlation function which will be used to derive the ensemble of non-Gaussian fields $\phi$ via the prescribed transformation $\phi=f(\varphi)$. For each realization, eq. \eqref{eq_1.1} is solved and a trajectory is obtained. The transport coefficients, diffusion and average velocity, are computed as simple statistical averages over the ensemble. DNS tries to mimic the whole (real) statistical problem resorting to numerical tools \cite{Palade2021,Vlad_2021,doi:10.1063/5.0035541}. 

The main source of errors in DNS is the insufficient numerical representation of the ensemble. In practice, we use a spectral representation of GRFs as discussed in \cite{Palade2021} with improved Eulerian and Lagrangian convergences:
\begin{align}\label{eq_1.8}
	\varphi(\mathbf{x}) = \sum_j^{N_c} S^{1/2}(\mathbf{k}_j)sin(\mathbf{k}_j\mathbf{x}+\frac{\pi}{4}\zeta_j)
\end{align}
where $\mathbf{k}_j$ are randomly distributed within the compact support of the spectrum $S(\mathbf{k})$ and $\zeta_j = \pm 1$ randomly chosen. In practice, we use $N_c\sim 10^d$ as it was found to ensure both the Gaussianity of the field as well as the details of the correlation function. Good statistical convergence is found both at Eulerian and Lagrangian level to be satisfied by ensembles with dimension $M\sim 10^5$. 

This is the standard approach of DNS on transport. It is suitable for GRFs since the field $\varphi$ is naturally Gaussian via the Central Limit Theorem. In order to tackle the problem of non-Gaussian $\phi$, we compute as such \eqref{eq_1.8} the fields $\varphi$ and solve the following eqn. of motion:
\begin{align}\label{eq_1.9}
	\frac{d\mathbf{x}(t)}{dt} = f^\prime (\varphi(\mathbf{x}(t),t))\hat{e}_z\times\nabla\varphi(\mathbf{x}(t),t) + \mathbf{V}_p
\end{align}

Note that the above equation \eqref{eq_1.9} is equivalent with eq. \eqref{eq_1.1} given the fact that $\phi = f(\varphi)$. The rest of the method remains unchanged. For more details see \cite{hauff_2006,Palade2021,Vlad_2021,doi:10.1063/5.0035541}.

\section{Results}
\label{section_3}

In order to capture the basic physical processes related to non-Gaussianity, we use two simple model correlation functions $\mathcal{E}_1$ and $\mathcal{E}_2$ for the $\varphi$ field:

\begin{align}\label{eq_2.1}
	\mathcal{E}_1(x,y,t)&=e^{-\frac{x^2}{2\lambda_x^2}-\frac{y^2}{2\lambda_y^2}-\frac{t}{\tau_c}}\\
	\mathcal{E}_2(x,y,t)&= \left(1+\frac{x^2}{\lambda_x^2}+\frac{y^2}{\lambda_y^2}\right)^{-1}e^{-t/\tau_c}
\end{align}
with $\lambda_x=\lambda_y=1$ and $\tau_c = 10$. Thus, $V_0  = 1$. These choices are in agreement with the gross features of turbulence spectra from incompresible plasma and fluids \cite{Boldyrev_2005,Gao_2015,Levinson_1984,PhysRevLett.102.165005}. 

\begin{figure*}
	\subfloat[\label{fig_1a}]{%
		\includegraphics[width=.49\linewidth]{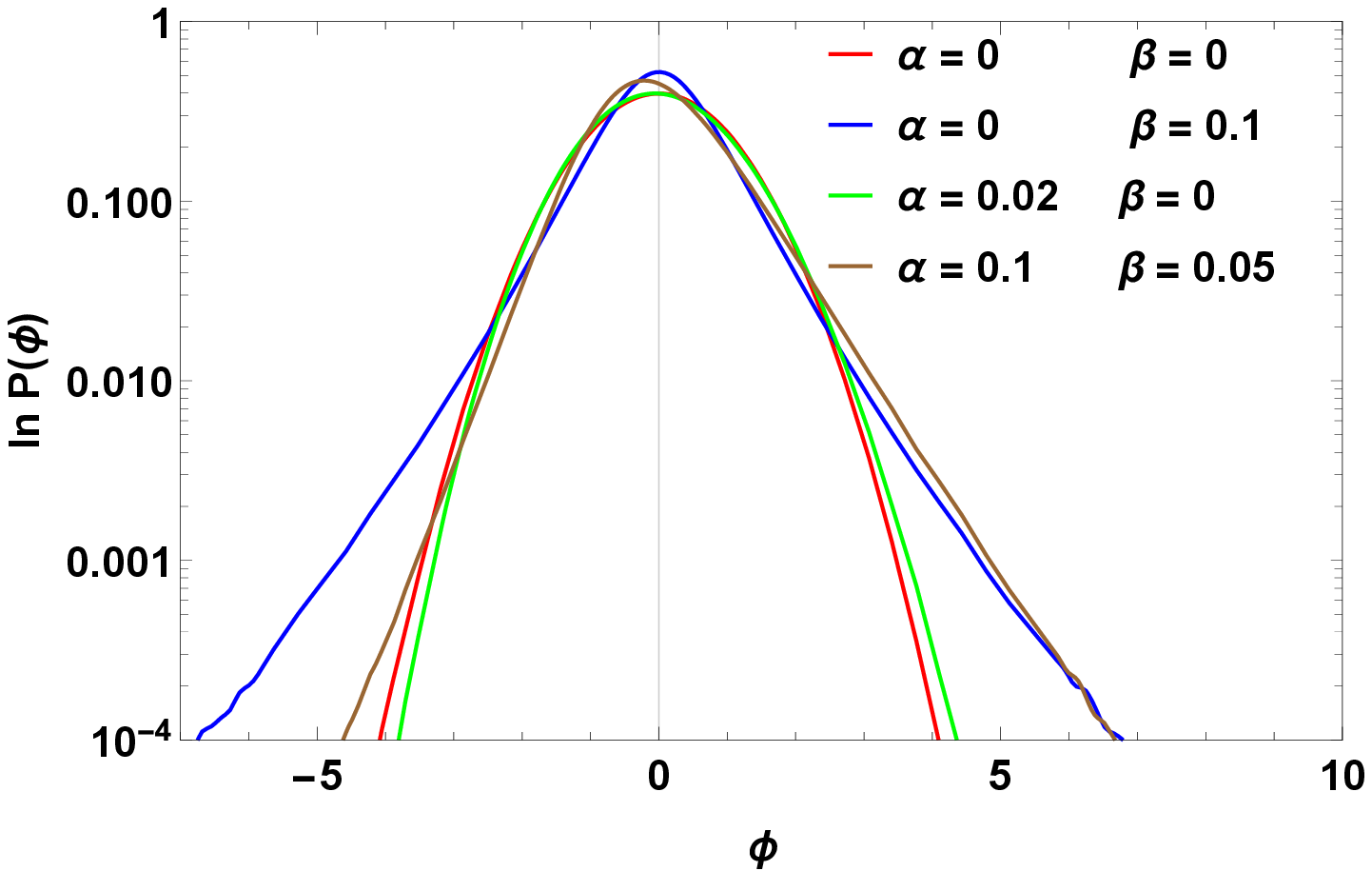}%
	}\hfill	\subfloat[\label{fig_1b}]{%
		\includegraphics[width=.49\linewidth]{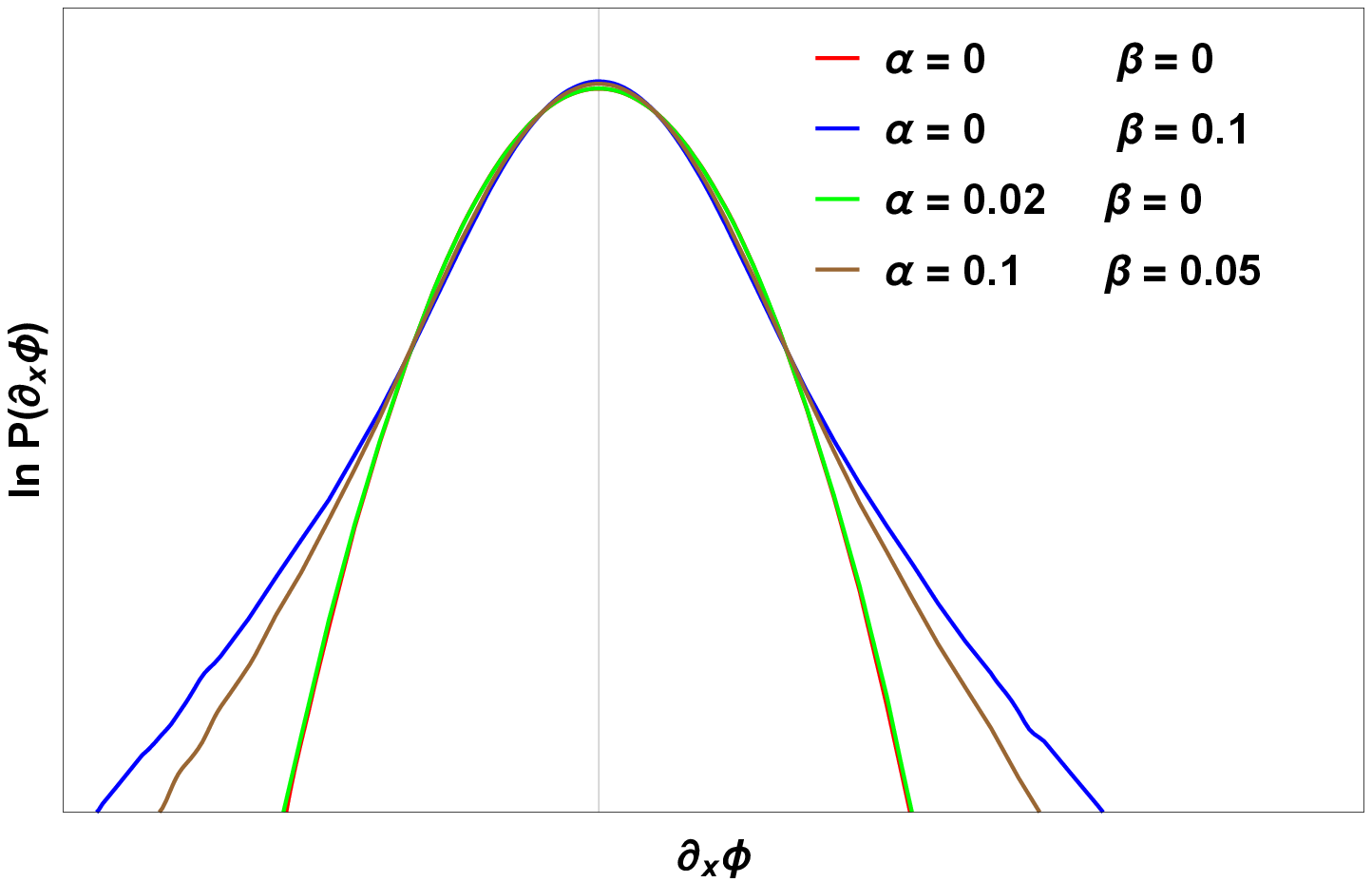}%
	}
	
	\caption{PDF of potential values (left) and potential derivatives (right) generated randomly in accordance with the non-linear transformation from eq. \eqref{eq_1.4}.}
\end{figure*}

In Fig. \ref{fig_1a},\ref{fig_1b} we show how the proposed transformation $f(\varphi) \sim \varphi+\alpha\varphi^2+\beta\varphi^3-\alpha$ distorts the Gaussian distribution both for the potential and its derivatives. Note how, through appropriate combinations of $\alpha$ and $\beta$ (the brown line), the resulting PDF is closer to Gaussianity on the negative domain $\phi<0$ and similar to an exponential distribution in the positive part $\phi>0$ (as observed in measurements). Also, due to the relation between $\partial_i\phi$ and $\varphi$, the distribution of derivatives $P(\partial_i\phi)$ is free of any skewness.

\begin{figure*}
	\subfloat[\label{fig_2a}]{
		\includegraphics[width=.49\linewidth]{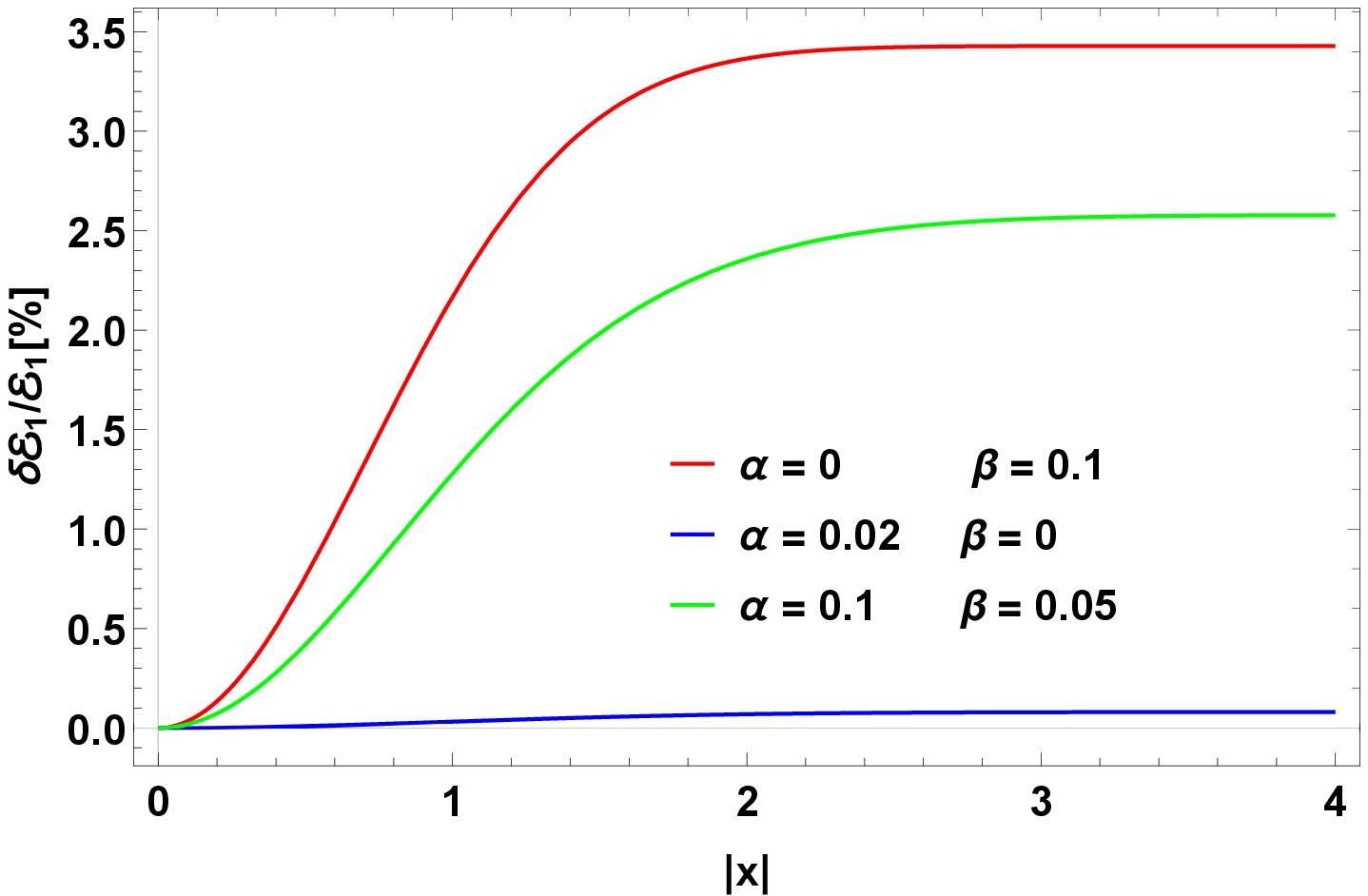}%
	}	
	\subfloat[\label{fig_2b}]{%
	\includegraphics[width=.49\linewidth]{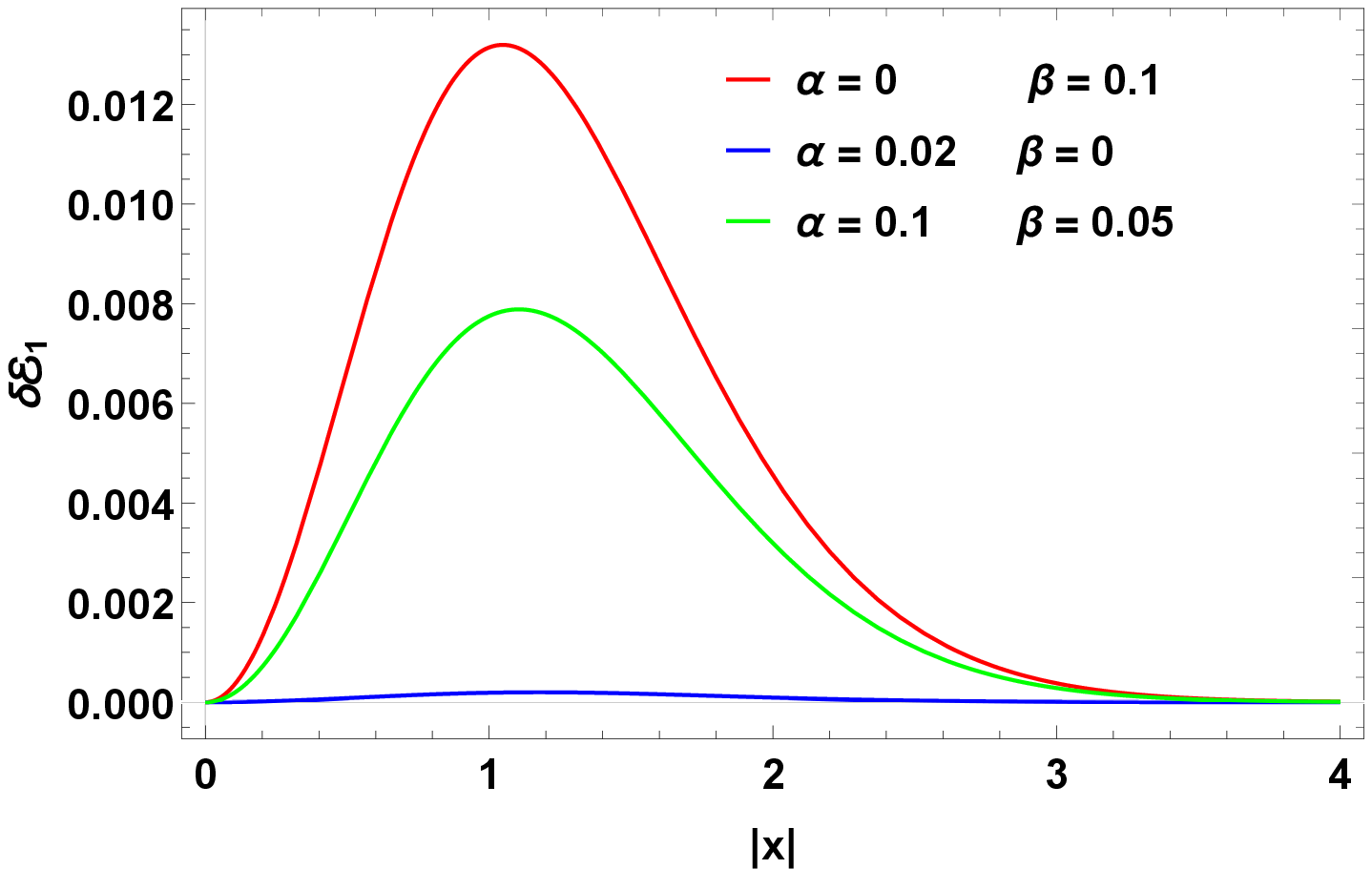}%
}	
	\caption{Relative changes $\delta \mathcal{E}_1 = \mathcal{E}_1^\prime - \mathcal{E}_1$ of the correlation function on the spatial domain at different small $\alpha, \beta$ values.}
\end{figure*}

In Figs. \ref{fig_2a},\ref{fig_2b} we plot the change in correlation $\delta \mathcal{E} = \mathcal{E}^\prime-\mathcal{E}$ under the effects of non-Gaussianity for the first model, $\mathcal{E}_1$. The results are in agreement with the analytical estimation \eqref{eq_1.4} that the departure is $\delta \mathcal{E}\sim\mathcal{O}(\alpha^2,\beta^2)\sim 1\%\mathcal{E}$ and virtually negligible, especially in the strong correlated area $|\mathbf{x}|\sim 0$. For $\mathcal{E}_2$ the profiles are extremely similar.

\subsection{Analytical estimations}
\label{section_3.1}

Our first level of analysis is the analytical one. We intend to estimate the change of the diffusion coefficient induced by non-Gaussian features of turbulence. In order to do that, we start by considering a simplified case of frozen turbulence, when the potential is time independent $\varphi(\mathbf{x},t) \equiv \varphi(\mathbf{x})$. This case can be obtained setting $\tau_c\to\infty$. Due to the Hamiltonian structure of the eqns. \eqref{eq_1.1}, the trajectories are closed and both the fictitious and the real potentials are conserved $\varphi(\mathbf{x}(t)) = \varphi(\mathbf{x}(0)) = \varphi(\mathbf{0})\implies\phi(\mathbf{x}(t)) = \phi(\mathbf{x}(0)) = \phi(\mathbf{0})$. For further purposes, let us denote the following function:

$$A[\varphi]=\frac{1+2\alpha \varphi+3\beta \varphi^2}{\sqrt{1+2\alpha^2+3\beta (2+5\beta)}}$$

With this notation, it turns out that the eqns. of motion in the Gaussian and non-Gaussian ($0$ index) cases become:

\begin{align}\label{eq_2.2}
	\frac{d\mathbf{x}_0(t)}{dt} &= \hat{e}_z\times\nabla\varphi(\mathbf{x}_0(t))\\
	\frac{d\mathbf{x}(t)}{dt} &= A[\varphi(0)]\hat{e}_z\times\nabla\varphi(\mathbf{x}(t))
\end{align}

It can be easily shown, via a variable transformation, that $\mathbf{x}(t) = \mathbf{x}_0(A[\varphi(\mathbf{0})]t)$. This exact relation allows us to relate the running diffusion coefficients between these two cases, non-Gaussian $D(t)$ and the Gaussian limit $D_0(t)$. In order to do that, let us denote the "conditional diffusion" for trajectories starting at equal potential values as: 

$$d_0(t;\varphi(\mathbf{0})) = \frac{1}{2}\frac{d}{dt}\langle \mathbf{x}_0^2(t) \rangle_{\varphi(\mathbf{0})}$$
for which it holds true that $d(t;\varphi(\mathbf{0}))= A[\varphi(\mathbf{0})]d_0(A[\varphi(\mathbf{0})]t;\varphi(\mathbf{0}))$. Finally, we write down: 

\begin{align}\label{eq_2.3}
	D_0(t) &= \int d\varphi(0) P[\varphi(\mathbf{0})]d_0(t;\varphi(\mathbf{0}))\\
	D(t) &= \int d\varphi(0) P[\varphi(\mathbf{0})]d_0(A[\varphi(\mathbf{0})]t;\varphi(\mathbf{0}))A[\varphi(\mathbf{0})]
\end{align}

Without any proof, we assume that some sort of \emph{generalized mean value theorem} is valid and these two integrals can be related through an effective potential:

\begin{align}
	\label{eq_2.4}
D(t) = A[\varphi_{eff}(t)] D_0(A[\varphi_{eff}(t)]t)
\end{align}


On the other hand, the anomalous feature of transport is reflected in the asymptotic behavior of running Lagrangian averages $L(t)$ as algebraic decays $L_0(t)\sim t^{-\gamma}$ \cite{Ottaviani_1992,RevModPhys.64.961,REUSS199894,Vlad_2004,PhysRevE.54.1857,PhysRevE.58.7359}. At small times $t\ll \tau_{fl}$ the dependence of $L(t)$ can be, usually, analytically computed. The presence of a decorrelation mechanism (finite $\tau_c$ in our case) tends to saturate asymptotically all Lagrangian quantities at values which can be estimated \cite{Vlad_2004,PhysRevE.58.7359} as $\lim_{t\to\infty} L(t) =L^\infty\approx L(\tau_c)$. We assume this to be true both for diffusion $D(t)$ and $\varphi_{eff}(t)$. 
	
Combining all these behaviors, the algebraic decay $D(t)\sim t^{-\gamma}, \varphi_{eff}(t)\sim t^{-\zeta}$, the approximate saturation at the decorrelation time $D^\infty \approx D(\tau_c), \varphi^\infty_{eff} \approx \varphi_{eff}(\tau_c)$ and the assumed relation between Gaussian and non-Gaussian diffusion \eqref{eq_2.4}, one can show that:

\begin{align}\label{eq_2.5}
	\begin{cases}
D^\infty/D_0^\infty = A[\varphi_{eff}(\tau_c;\alpha,\beta)]^2 \approx 1 + 2 \alpha^2 + 12 \beta^2,    	&K_\star\ll 1\\	
D^\infty/D_0^\infty = A[\varphi_{eff}(\tau_c;\alpha,\beta)]^{1-\gamma} \approx 1 +3 \beta(-1 + \gamma), 		&K_\star\gg 1
	\end{cases}
\end{align}

These estimations suggest that $\alpha$ has only a quadratic effect on the diffusion coefficient. This aspect can be understood from another perspective, analyzing how the Lagrangian correlation of velocities $L_v(t) = \langle v_x(0)v_x(t)\rangle$ (the time derivative of $D(t)$) varies with $\alpha$ up to the first order:

$$\frac{\partial }{\partial\alpha}L_v(t) = \frac{\partial}{\partial\alpha}\langle v_x(\mathbf{0},0)v_x(\mathbf{x}(t),t)\rangle \propto  \langle \varphi(0)\partial_y\varphi(0)\partial_y\varphi(t)\rangle+\langle \varphi(t)\partial_y\varphi(0)\partial_y\varphi(t)\rangle \approx 0$$

Note that $\varphi(0)$ and $\partial_i\varphi(0)$ are uncorrelated Gaussian quantities. Moreover, Lumley's theorem \cite{Monin} assures us that the space derivatives $\partial_i\varphi(t)$ remain Gaussian quantities at all times. Only the distribution of Lagrangian potentials $\varphi(t)$ might depart from Gaussianity in case of finite $\tau_c$, but only slightly. Thus, the derivative $\partial_\alpha L_v(t)$ is roughly made up of averages of products of three Gaussian quantities, therefore, is zero. This means that  $L_v(t)$ is roughly independent of $\alpha$ up to first order. The same goes for the diffusion.

Following the above reasoning and estimations \eqref{eq_2.5}, we expect that the non-Gaussian diffusion will vary roughly as $\mathcal{O}(\beta), \mathcal{O}(\alpha^2)$. For this reason, we define a response function (susceptibility $\chi$) to quantify the possible linear dependency between diffusion variation and turbulence excess kurtosis (as a measurable quantity):

\begin{align}\label{eq_2.6}
	\chi = \lim_{\delta\kappa\to 0 }\frac{1}{\delta\kappa}\left(1-\frac{D^\infty(\delta\kappa)}{D^\infty(0)}\right)
\end{align}

\subsection{Numerical results}
\label{section_3.2}

We intend to test further if the analytical estimations found above bear any meaning in real situations. For that, we use the statistical methods described previously: DNS and DTM. We underline that DNS is an exact-in-principle method which is hindered in practice only by the numerical resolution, thus, it requires a large amount of CPU resources. DTM is an approximation which provides only-qualitative results and it is easy to implement numerically. The purpose of DTM in this work is to serve as a supplementary test for the results of DNS which might be plagued with a small, but uncertain, degree of numerical inaccuracies.
\begin{figure}
	\centering
	\includegraphics[width=0.7\linewidth]{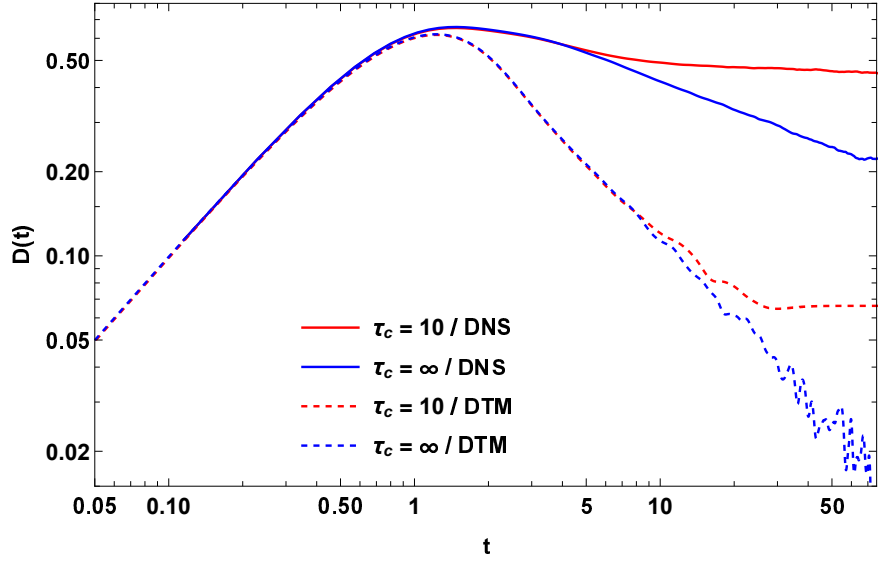}
	\caption{Running diffusion coefficient $D(t)$ obtained for the correlation $\mathcal{E}_1, V_p = 0$ in the case $\tau_c\to\infty$ (blue) and $\tau_c = 10$ (red) with the use of DNS (full line) and DTM (dashed line).}
	\label{fig:fig3}
\end{figure}

We perform numerical simulations of the running diffusion coefficients $D(t)$ for the incompressible motion \eqref{eq_1.1} where the potential $\phi(\mathbf{x},t)$ is described via a fictitious field $\varphi(\mathbf{x},t)$ \eqref{eq_1.4} with known Eulerian correlation \eqref{eq_2.1}. The numerical method for trajectory propagation both in \eqref{eq_1.9} (DNS) and \eqref{eq_1.5a} (DTM) is a 4th order Runge-Kutta method with a fixed time-step $\Delta t\sim 10^{-1}min(\tau_c,\lambda_c^2)$. The simulation time is $t_{max}\sim 5\tau_c$. The number of trajectories simulated with DNS is routinely $N_p \sim 10^5$ while the number of subensembles used in DTM $N_s\sim 10^5$. These resolutions are chosen for numerical accuracy and statistical precision. Using dedicated programming procedures, typical simulations on personal computers require in terms of CPU time: $t_{CPU}^{DTM}\sim 1min$ and $t_{CPU}^{DNS}\sim 10min$. 

\begin{figure}
	\centering
	\includegraphics[width=0.7\linewidth]{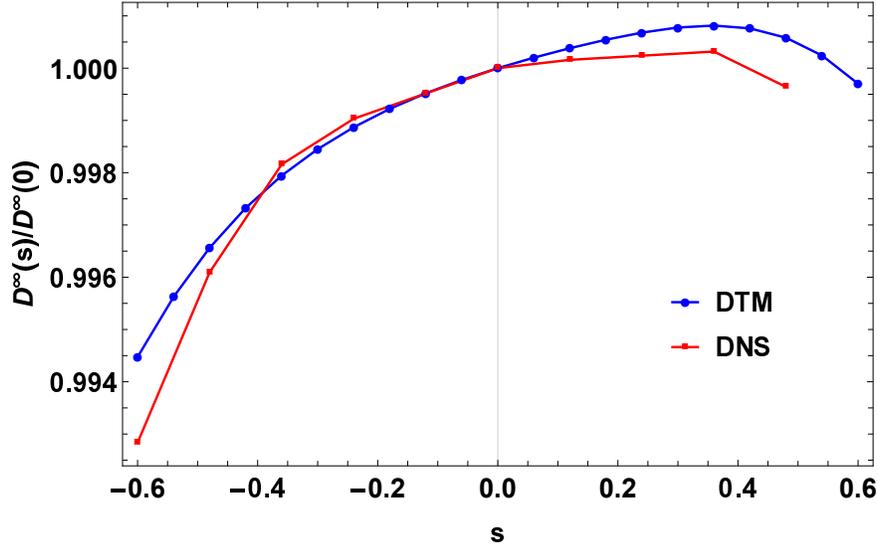}
	\caption{Asymptotic values of diffusion at $\beta = 0, \tau_c = 10, V_p = 0$ vs. the skewness $s$ obtained with the use of DTM (blue line) and DNS (red line). }
	\label{fig:fig4}
\end{figure}

Beyond diving into the matter of non-Gaussianity, let us have a look at a typical running diffusion coefficient $D(t)$ obtained in the case $\alpha = \beta = 0$. The results of both methods are shown in Fig. \ref{fig:fig3} at $\tau_c = 10$ and $\tau_c\to\infty$. In the case of frozen turbulence one can see the algebraic decay of diffusion $D(t)\sim t^{-\gamma}$. The effect of finite $\tau_c$ is the saturation of diffusion to a constant value $D^\infty = D(t\to\infty)$. Note that DTM reproduces the qualitative behavior of trapping and deccorelation at all times. Yet, the results are quantitatively different in the asymptotic region $t\gg\tau_{fl}$. This is a due to the overestimation of trapping in the DTM approximation. Consequently, DTM overestimates the $\gamma$ exponent too and we expect it will overestimate the effect of intermittency in the strong (low-frequency) turbulence regime. 
\begin{figure}
	\centering
	\includegraphics[width=0.7\linewidth]{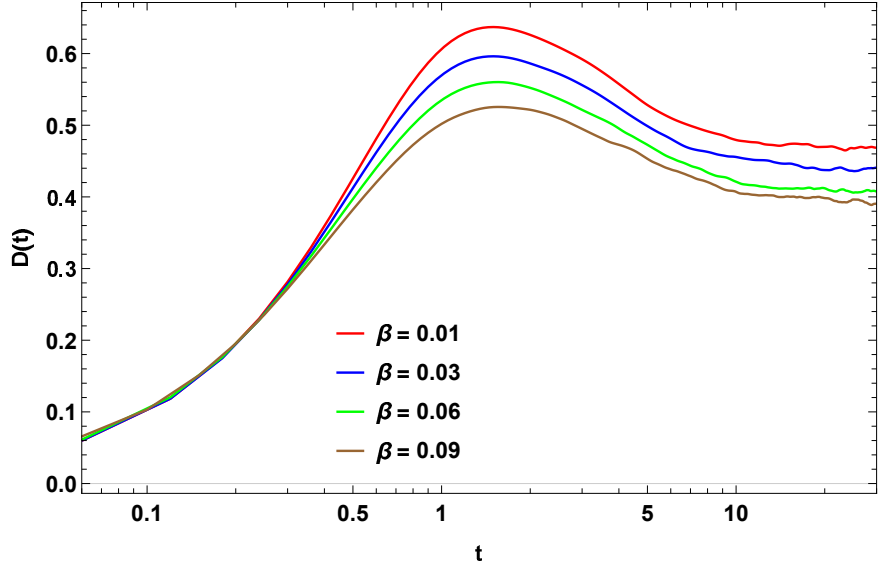}
	\caption{Running diffusion profiles $D(t)$ obtained with DNS for the correlation $\mathcal{E}_1$ with $\tau_c =10$ and $V_p = 0$ at different $\beta$ values and $\alpha = 0$.}
	\label{fig:fig5}
\end{figure}

We investigate further the dependence of the diffusion on $\alpha$ and find that both methods, DTM and DNS predict a negligible  variation (Fig. \ref{fig:fig4}). The results are in line with our analytical estimation that $\alpha$ affects the transport only second order. A very weak linear dependence supplemented by a weak quadratic one can be observed.

Given this fact, let us set $\alpha = 0$ and look further how $\beta$ affects the diffusion. The mechanism can be seen at work in Fig. \ref{fig:fig5} where several running diffusion profiles $D(t)$ are shown for different $\beta$ values. As expected, the effect is visible only at larger times, at least at the order of $\tau_{fl}$ and it results in a decrease of the diffusion coefficient. This behavior can be quantified further by inspecting the variation of the asymptotic diffusion $D^\infty$ with $\beta$. 
\begin{figure}
	\centering
	\includegraphics[width=0.7\linewidth]{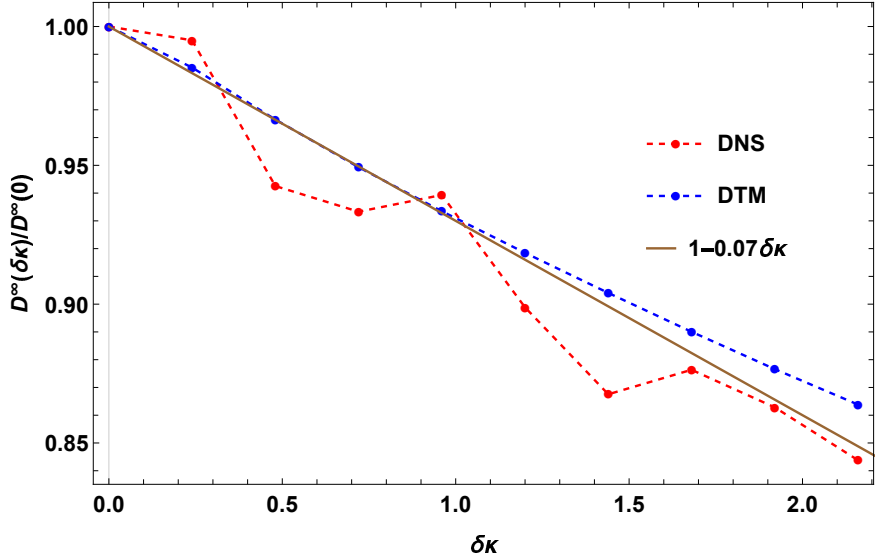}
	\caption{The relative asymptotic diffusion dependence with the excess kurtosis obtained with DTM for $\mathcal{E}_2$ (blue line) and with DNS for $\mathcal{E}_1$ (red line).}
	\label{fig:fig6}
\end{figure}

We plot in Fig. \ref{fig:fig6} asymptotic diffusion coefficients computed for different values of $\beta$ with DNS (blue) for the correlation function $\mathcal{E}_1$ and with DTM (red) for $\mathcal{E}_2$. Our expectation that the $\beta$ parameter drives a linear change in the diffusion is confirmed. Also, it must be emphasized how close are the results of DTM to those of DNS, given the fact that they use two distinct correlations.

\begin{figure}
	\centering
	\includegraphics[width=0.7\linewidth]{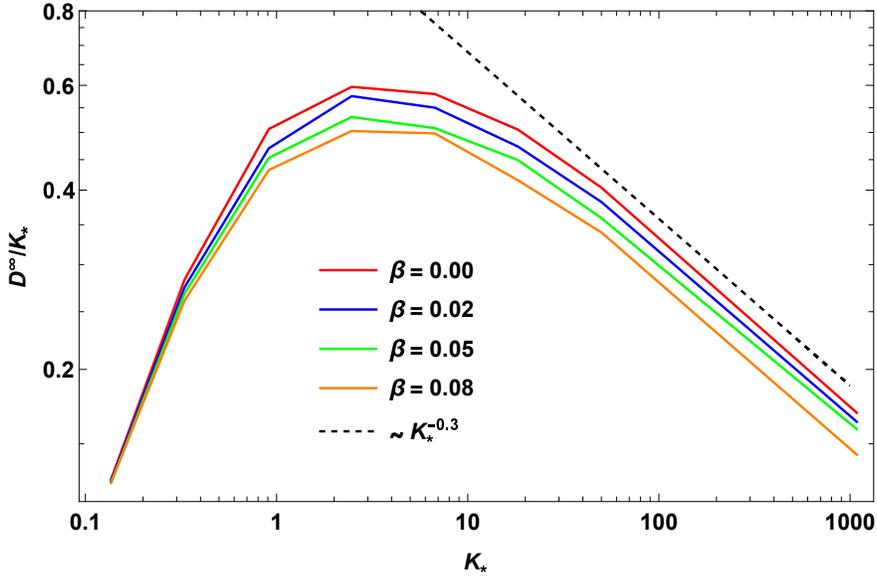}
	\caption{Asymptotic diffusion coefficient vs. the Kubo number at different $\beta$ values. The results are obtained with DNS for $\mathcal{E}_1$ and $V_p = 0$.}
	\label{fig:fig6a}
\end{figure}

Going further into understanding the effects of intermittency, we plot in Fig. \ref{fig:fig6a} profiles of asymptotic diffusion coefficients vs. the Kubo number at different values of the $\beta$ parameter. As expected from the results of Fig. \ref{fig:fig6} the effect of $\beta$ is to inhibit the overall values of diffusion. Yet, something supplementary must be underlined: in the high $K_\star$ region, that of trajectory trapping, the profiles are approximate parallel to each other and to the $\sim K_\star^{-0.3}$ line. This tells us that the Kubo no. scaling is universal $\gamma = 0.3$ and it is unaffected by non-Gaussianity.

\begin{figure}
	\centering
	\includegraphics[width=0.7\linewidth]{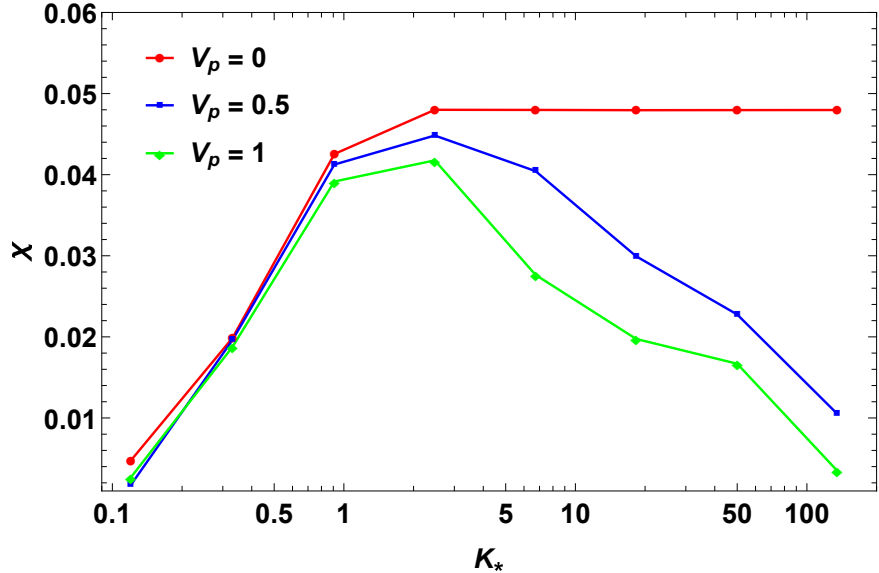}
	\caption{Susceptibility $\chi$ as a function of Kubo number $K_\star$ obtained with DNS for $\mathcal{E}_1$ at distinct $V_p$ values.}
	\label{fig:fig7}
\end{figure}

After confirming the linear behavior in $\beta$, we look into the dependence of the susceptibility $\chi$ versus the Kubo number $K_\star$. We have performed extensive numerical simulations both with DTM and DNS varying both the Kubo number (through $\tau_c$) and the $\beta$ parameter. Final result are shown in Fig. \ref{fig:fig7} where we plot $\chi(K_\star)$ obtained with DNS at several distinct $V_p$ values. As one can see, at small correlation times, this quantity is null. It increases only quadratically with $K_\star$, until around $\tau_c$ of the order of the time of flight and saturates for $V_d=0$. An interesting effect of the average velocity can be seen in Fig. \eqref{fig:fig7}, which consists in the strong attenuation of the effect of the kurtosis. The susceptibility has a strong algebraic decay in these conditions.

\begin{figure}
	\centering
	\includegraphics[width=0.7\linewidth]{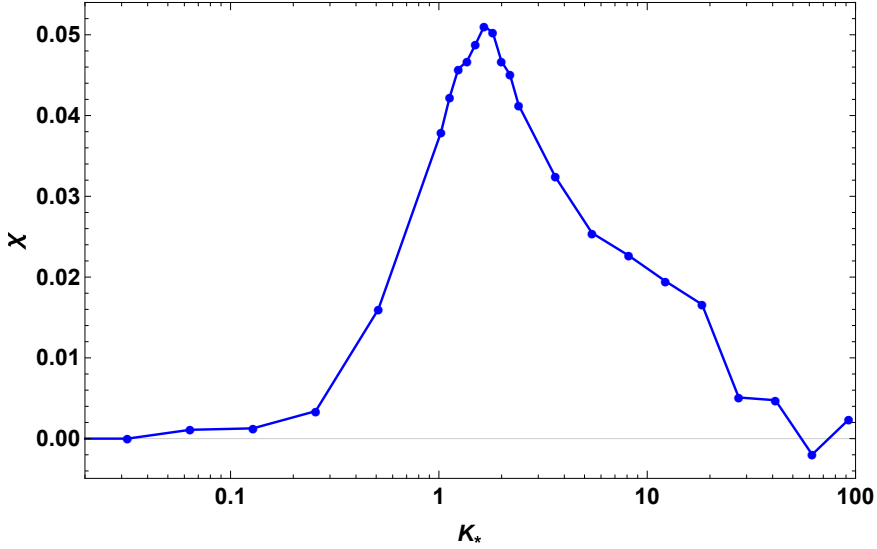}
	\caption{Susceptibility $\chi$ as a function of Kubo number $K_\star$ obtained with DTM for $\mathcal{E}_2$ at $V_p = 0$.}
	\label{fig:fig7a}
\end{figure}

In Fig. \eqref{fig:fig7a} we show the profile of $\chi$ obtained with DTM in the case of $V_p = 0$. The method is able to confirm what was found with DNS: the susceptibility grows up to a maxima around the time-of-flight and decays at larger values of $K_\star$. As expected, since DTM overestimates the trapping, it also overestimates the decay of $\chi$.

\subsection{Microscopic analysis}
\label{section_3.3}

The effect of non-Gaussianity on transport can be understood from a microscopic perspective, following how individual trajectories change, or how their statistics is modified. 

\begin{figure*}
\label{fig_8}
		\includegraphics[width=.77\linewidth]{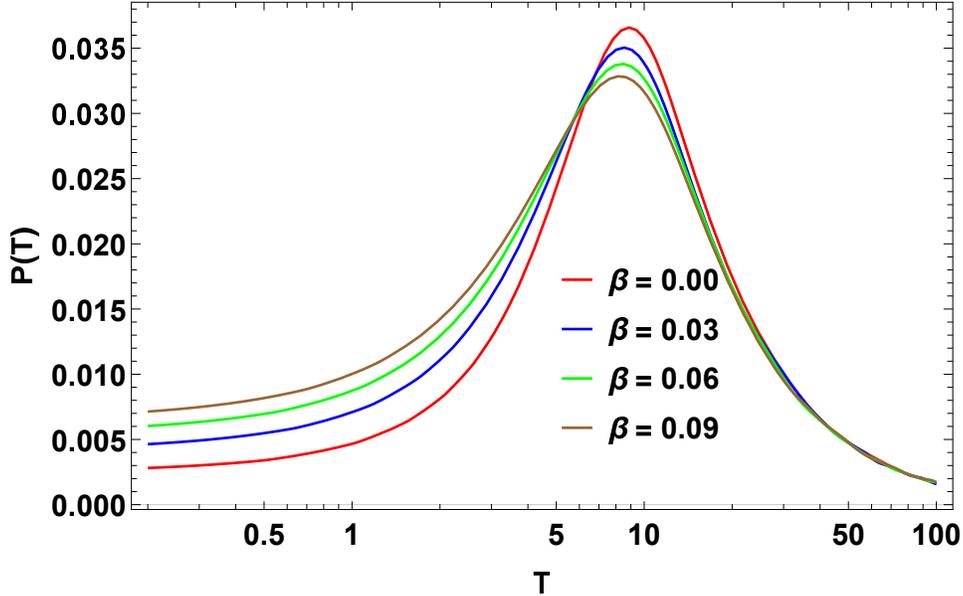}%
	\caption{PDF $P(T)$ of the trajectory periods at different $\beta$ values in frozen turbulence $\tau_c\to \infty$ and no poloidal velocity $V_p = 0$.}
\end{figure*}

One can start the analysis from the limiting case of frozen turbulence $\tau_c\to\infty$. While the trajectories remain unchanged (see Section \ref{section_3.1}) the velocity is changed with a factor $A[\varphi(\mathbf{0})]$. The consequence is that the diffusion across that particular equipotential line becomes $d(t;\varphi(\mathbf{0})) = A[\varphi(\mathbf{0})]d_0(A[\varphi(\mathbf{0})]t;\varphi(\mathbf{0}))$. This is equivalent with a change of trajectory's period by a factor $A^{-1}[\varphi(\mathbf{0})]$. The statistical effects can be seen in Figs. \ref{fig_8} where the PDF of the periods $P(T)$ is plotted at different $\beta$ values with $\alpha = 0$. 

One can notice that the intermittency lowers the general values of the periods. This is a natural consequence of the fact that $A[\varphi(0)]>1$. The effect is more pronounced at small and intermediate values $T\sim\tau_{fl}$ since the low-frequency trajectories, i.e. $T\gg \tau_{fl}$, are those with low values of the potential $\varphi(\mathbf{0})\sim 0$. For the latter, the factor $A[\varphi(\mathbf{0})]\sim 1$. 

Although at small times $t\ll \tau_{fl}$ all trajectories contribute to the diffusion, their average is only slightly dependent on $\beta$. The effect becomes more pronounced at times at the order of the time of flight. The fact that the distribution of slow trajectories $T\gg \tau_{fl}$ are almost unchanged explains why the asymptotic behavior of diffusion in the non-linear regime $K\gg 1$ is universal: the scaling law $D\sim K^{1-\gamma}$ is invariant to non-Gaussianity at $K\gg 1$. 

On the other hand, the distribution of potentials is changed. Thus, the effects of non-Gaussianity on diffusion results both from a change in the weight of each equipotential line and from the distortion of time periods.

Furthermore, we underline that, if the turbulence is frozen, both the distribution of Lagrangian potentials and that of Lagrangian velocities (Lumley's theorem, due to the divergenceless property of the Eulerian velocity field) are invariant in time. Under these two strong constrains, it is clear that the only microscopic effect of non-Gaussianity is the redistribution of trajectory's frequencies. Other dynamical phenomena which might affect the transport are not present.

Finally, we note how the distribution of small valued potentials $\phi\ll \Phi$ is virtually unchanged due tot the shape of $f$ (the non-linear mapping between $\varphi$ and $\phi$). These small values are linked to long -low frequency- trajectories, thus, to the behavior of diffusion in the non-linear regime $K\gg 1, \tau_c\gg\tau_{fl}$. This explains why the scaling behavior ($\gamma$) is unchanged by intermittency.

\section{Conclusions}
\label{section_4}

In the present work we have analyzed the effects of intermittency on turbulent transport in magnetized fusion plasmas. The intermittent phenomena are included in the turbulent electric potential $\phi(\mathbf{x},t)$ as non-Gaussian features of its distribution in agreement with experimental data. We consider the simple case of $\mathbf{E}\times\mathbf{B}$ drift-type dynamics in a constant magnetic field in slab geometry. The transport is characterized by the resulting diffusion coefficient.

In order to mimic the experimental data, the non-Gaussian field $\phi$ is modeled using a non-linear transformation $f$ from a fictitious Gaussian turbulent field $\varphi$. The transport and the turbulence model are analyzed on three distinct levels: analytical, numerical and physical. 

The analytical analysis suggests that, for small values of skewness and kurtosis, the diffusion decreases linearly with the excess kurtosis $\delta \kappa$ while the dependency on skewness can be neglected. The numerical analysis, which is performed using two distinct statistical methods (DTM and DNS), confirms the analytical estimations: indeed, only the kurtosis of the non-Gaussian field affects the diffusion in a linear manner. Moreover, exploring numerically the main dependence of the response coefficient $\chi(K_\star)$ on correlation time, it was found an interesting behavior. In the quasilinear regime, the effects of intermittency are small $\chi \to 0$. At the other end of the spectrum, $K_\star \gg 1$, $\chi$ saturates to a maxima which is reached after the time of flight $\tau_{fl}$, i.e. $K_\star = 1-2$. The presence of a poloidal velocity induces an algebraic decay of the susceptibility $\chi \sim K_\star^{-\zeta}\to 0$ in the non-linear regime. To summarize our findings: 

\begin{align}
\begin{cases}	D^\infty(\delta \kappa,s)\approx D^\infty(0,0)\left(1+\chi(K_\star)\delta \kappa \right)&\\
	\chi(K_\star) \propto K_\star^2 , &K_\star\ll \tau_{fl}\\
	\chi(K_\star) \sim  K_\star^{-\zeta} , &K_\star\gg \tau_{fl}\\	
	max[\chi(K_\star)] \approx \chi(\tau_{fl} = \tau_c)
\end{cases}
\end{align}

Our results suggest that the specific correlation time of turbulence $\tau_c$ as well as the departure from Gaussianity might serve as control parameters for the anomalous transport of plasma. 

\section*{Acknowledgement}

This work has been carried out within the framework of the EUROfusion Consortium and has received funding from the Euratom research and training programme 2014-2018 and $2019-2020$ under grant agreement No $633053$ and from the Romanian Ministry of Research and Innovation. The views and opinions expressed herein do not necessarily reflect those of the European Commission.

\section*{Data availability}

The data that support the findings of this study are available from the corresponding author upon reasonable request.

\bibliographystyle{unsrt}
\bibliography{biblio}

\end{document}